 \documentclass[final,3p,times,twocolumn]{elsarticle}

\usepackage{amssymb}
\usepackage{graphicx}
\usepackage{color}

\journal{Thermochimica Acta}

\begin{document}

\begin{frontmatter}



\title {Experimental evaluation and thermodynamic assessment of the LiF$-$LuF$_3$ phase diagram}



\author[IPEN]{I. A. dos Santos\corref{cor1}}
\cortext[cor1]{corresponding author}
\ead{iasantosif@usp.com.br}

\author[IKZ]{D. Klimm}
\author[IPEN]{S. L. Baldochi}
\author[IPEN]{I. M. Ranieri}

\address[IPEN]{Instituto de Pesquisas Energ\'eticas e Nucleares, CP 11049, Butant\~a 05422-970, S\~ao Paulo, SP, Brazil}
\address[IKZ]{ Leibniz Institute for Crystal Growth, Max-Born-Stra\ss e 2, 12489 Berlin, Germany}

\begin{abstract}
The phase diagram of the system LiF$-$LuF$_3$ has been revised using thermal analysis. Specific heat capacity and enthalpy of phase transition and fusion were measured by differential scanning calorimetry for all compounds belonging to the system. A thermodynamic optimization of the LiF$-$LuF$_3$ phase diagram was performed by fitting the Gibbs energy functions to experimental data that were taken from the literature or measured in this work. Excess energy terms, which describe the effect of interaction between the two fluoride compounds in the liquid solution, were expressed by the Redlich-Kister polynomial function. The theoretical phase diagram assessed was in suitable agreement with the re-evaluated experimental data. 
\end{abstract}


\end{frontmatter}


\section{Introduction}
\label{sec:intro}

LiLuF$_4$ (LLF) crystals have been largely investigated for doping with rare earth ions, mainly focusing on the development of laser media \cite{librantz00, librantz01, Preussler11, Chen11, Tonelli00}. Taking into account that LLF has the most compact crystalline structure among the LiREF$_4$ (RE= rare earth element or Y, respectively) crystals family, some studies  have also reported the Li(Lu,RE)F$_4$ (RE = Gd or Y) mixed crystals as an alternative to achieve better optical properties, either wider emission bandwidths \cite{iasantos00, Vieira02} or  improvement of their photochemical stability \cite{Nizamutdinov} for rare earth doped  crystals.

Conversely, there is a lack of information concerning the thermodynamic properties of LLF and the corresponding LiF$-$LuF$_3$ phase diagram. Heat capacity and enthalpy data for LuF$_3$ were initially investigated by Spedding \cite{Spedding71} and, more recently, Lyapunov \textit{et al.} revised the enthalpy data and proposed a new $C_P(T)$ function for this compound \cite{Lyapunov04}. No thermodynamic data are available for LLF whilst LiF data are given by the Barin compilation \cite{Barin93}.

The phase diagram of the binary system LiF$-$LuF$_3$ was initially studied in the 1960s by Thoma \textit{et al.} \cite{thoma01}. In that work, the LiF$-$LuF$_3$ phase diagram was described with an intermediary compound (LLF) which melts congruently at 1098\,K. Two eutectic points were settled, one at 22\,mol\% LuF$_3$ and 968\,K and the other at 54\,mol\% LuF$_3$ and 1083\,K. The polymorphic transition from orthorhombic to hexagonal structure in LuF$_3$ was reported at 1218\,K. Harris \textit{et al.} \cite{Harris83} revised this phase diagram in the 1980s, determining the melting point of LLF at 1123\,K, the first eutectic was reported at 20\,mol\% LuF$_3$ and 977\,K and the second one at 58\,mol\% LuF$_3$ and 1105\,K. Both of them used thermal analysis to determine the invariant reactions and respective temperatures. In this work, the LiF$-$LuF$_3$ binary system was experimentally revised through DSC technique. The enthalpy of phase transition and fusion for all compounds which belong to this system were calculated. Using the obtained data and those collected from literature \cite{Lyapunov04, Barin93}, LiF$-$LuF$_3$ theoretical phase diagram has been optimized. Gibbs excess energy terms for the liquid solution, which describe the effects of interaction between the two fluorides, were expressed by the Redlich-Kister polynomial function \cite{Redlich00}. Enthalpy of formation and entropy at 298.15\,K were assessed for LuF$_3$ and LLF.

\section{Experimental}
\label{sec:exp}

Samples with the desired proportions were mixed using commercial LiF (AC Materials, 99.999\%) and LuF$_3$ (AC Materials, 99.99\%), and afterwards were pulverized in a mortar. DSC curves were obtained using a Netzsch STA 409 PC Luxx heat-flux differential scanning calorimeter. The sample carrier was calibrated for $T$ and sensitivity at the phase transformation points of BaCO$_3$ and at the melting temperatures of In, Zn and Au. The experiments were carried out under Ar flow of 30\,cm$^3$/min, using Pt/Au crucibles with lid. Two heating/cooling cycles with heating rate of 10\,K/min and samples masses around 30\,mg were adopted. The melting point of compounds, the temperature of the solid phase transition and of the invariant reactions were calculated considering the extrapolated onset of the thermal event. The liquidus temperatures of the intermediary compositions were evaluated from the extrapolated offset temperatures.

The heat capacity measurements were performed using the same equipment described above, where a proper sample carrier for $C_P$ analysis was installed. The experiments consisted of three steps, an isothermal segment at 40$^{\,\circ}$C for 20\,min, a dynamic heating segment with heating rate of 10\,K/min and a final isothermal segment at maximum temperature for 5\,min. Other experimental conditions were kept as mentioned previously for simple DSC experiment. LuF$_3$ and LiLuF$_4$ heat capacity were determined according to ratio method. In this method, the sample DSC heat flow signal is compared to the DSC signal of a calibration standard of known specific heat (sapphire in this case). Both curves are corrected by a baseline correction experiment where empty reference and sample crucibles are placed in the DSC furnace and the system signal drift is measured under identical experimental conditions. 

\begin{table*}[ht]
\renewcommand\footnoterule{}
\caption{$\Delta H${(298.15\,K)} (kJ\,mol$^{-1}$), $S${(298.15\,K)} (J\,K$^{-1}$\,mol$^{-1}$), $\Delta H_{PT}$ (Phase transition: heat of fusion or polymorphic transition) (kJ\,mol$^{-1}$) and $C_P$ data $a,b,c$ entering eq. (\ref{eq1}) for LiF, LuF$_3$ and the intermediate compound LiLuF$_4$.}
	  \begin{minipage}{15cm}
	  \renewcommand{\arraystretch}{1.1}
	  \centering
	  \begin{tabular}{lrrrrrrrr}
		\hline
Compound   &  $\Delta H${(298.15\,$K$)}   &   $S${(298.15\,$K$)}    &  $\Delta H_{PT}$ (Lit) & $\Delta H_{PT}$\footnote{$\Delta H_{PT}$ and $C_P$ function were measured by the DSC technique; $\Delta H${(298.15\,$K$)} and $S${(298.15\,$K$)} were assessed in this work.}&    $a$      &      $b$     & $c$ \\
\hline
LiF$(S)$\footnote{Data taken from Barin \cite{Barin93}.}   &  $-616.931$   &    35.660   &   27.09   &   27.68  &  42.689   &    $1.742\times10^{-2}$ & $-5.301\times10^{5}$\\
LiF$(l)^b$   &  $-594.581$   &    42.997   &     --                   &     --    &  64.183   &      --        &      --     \\
LuF$_3(S1)^a$ &  $-1697.662$  &   118.420  &  25.40\footnote{Data taken from Lyapunov \textit{et al.} \cite{Lyapunov04}.}  &   18.28   & 97.496   & $9.460\times10^{-3}$ & $-9.998\times10^5$	\\
LuF$_3(S2)^a$ &  $-1693.056$  &   111.358   &   29.90$^c$  &   23.34   & 114.500$^c$  &  --    &  --   \\
LuF$_3(l)^a$   &  $-594.581$   &    42.997   &     --                   &     --    &  131.800$^c$   &      --        &      --     \\         
LiLuF$_4 ^a$   & $-2323.639$  &   153.218  &   --   &   63.538     &   139.430  &   $3.200\times10^{-2}$  &  $-1.881\times10^6$  \\
\hline
		\end{tabular}
		\end{minipage}
   	\label{table01}
	  \end{table*}

\section{Thermodynamic method}
\label{Thermo}

A thermodynamic simulation of a $T - X$ binary phase diagram requires the description of Gibbs energy functions for all compounds in the system and the Gibbs functions of mixing, if solution phases are present. In most cases the excess Gibbs functions are unknown for the solutions, therefore a thermodynamic assessment is necessary in order to determine thermal effects of mixing. Gibbs energy for a solid compound is defined as a function of enthalpy and entropy at the reference temperature state (298.15\,K) and can be obtained from the heat capacity function ($C_P(T)$) as follows:

\begin{equation}
C_P(T)=a+bT+cT^{-2}
\label{eq1} 
\end{equation}

\begin{equation}
H(T)=H^{0}_{298.15\,\mathrm{K}}+\int_{298.15\,\mathrm{K}}^{T}C_P \,\mathrm{d}T
\label{eq2} 
\end{equation}

\begin{equation}
S(T)=S^{0}_{298.15\,\mathrm{K}}+\int_{298.15\,\mathrm{K}}^{T}\frac{C_P}{T} \,\mathrm{d}T
\label{eq3} 
\end{equation}

Combining the equations (\ref{eq2}) and (\ref{eq3}), the $G(T)$ equation is given as follows:
\begin{eqnarray}
G(T)=H^{0}_{298.15\,\mathrm{K}}-(S^{0}_{298.15\,\mathrm{K}})T \nonumber\\
+\int_{298.15\,\mathrm{K}}^{T}C_P \,\mathrm{d}T - T\int_{298.15\,\mathrm{K}}^{T}\frac{C_P}{T} \,\mathrm{d}T
\label{eq4} 
\end{eqnarray}

Usually the heat capacity is the most accessible physical property to be measured. $C_P(T)$ functions can be determined by fitting a set of experimental data at a proper polynomial function (\textit{e.g.} equation \ref{eq1}). Depending on the compound, more terms are added or disregarded in the polynomial function to obtain the best fitting.  According to equation (\ref{eq1}) and considering the equations (\ref{eq2}) and (\ref{eq3}), the enthalpy of formation, the absolute entropy at the reference temperature and the heat capacity are required to determine the minimum of $G$ and thus the thermodynamic equilibrium. These data are not available for the LLF, therefore the experimental data calculated by DSC was assumed to set the $C_P$ equation, as will be discussed later. $C_P$ equation for LuF$_3$ was also obtained using this technique and was compared with the literature values. Enthalpy and entropy at 298.15\,K for LLF and LuF$_3$ were properly assessed by optimization. Table \ref{table01} summarizes all calorimetric data used on the phase diagram assessment.The thermodynamic data for the LiF end member is well established and was taken from Barin compilation \cite{Barin93}.

For the liquid solution phase, function $G(T)$ is expressed as the sum of the Gibbs energy weighed contribution of the pure compounds ($G_0$), the contribution of an ideal mixture ($G_{ID}$) and finally a term related to the non-ideal interaction, defined as the excess energy ($G_{ex}$). The sub-regular solution model of Redlich-Kister was adopted to describe the excess energy of the liquid phase in this system \cite{Redlich00}, and is given by:
\begin{equation}
G_{ex}=x_Ax_B\sum_{j=0}^{N}L_j\,(x_A-x_B)^j
\label{eq5} 
\end{equation}
where $x_A$ and $x_B$ are the molar fractions of components $A$ and $B$, respectively. $L_j$ terms represent the interaction coefficients between the basis compounds and they are given as a linear function of temperature. The optimization was performed using the OptiSage module in the FactSage 6.2 software \cite{FactSage6_2}, which uses the Bayesian Algorithm \cite{Pelikan00}. This algorithm is based on a probability model to obtain the fit between the theoretical Gibbs energy functions and the experimental data. 

\section{Results and discussion}
\label{sec:results}

\subsection{Experimental results}

\begin{figure}[htb]
\centering
\includegraphics[width=0.45\textwidth]{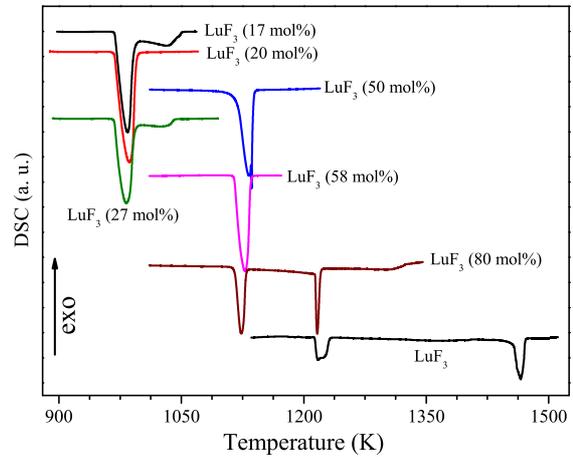}
\caption{Second heating DSC curves for some compositions in the LiF--LuF$_3$ system.}
\label{fig1}
\end{figure}

DSC curves for pure LuF$_3$ and six different compositions are shown in Figure~\ref{fig1}. The curves for 17, 20 and 27\,mol\%\,LuF$_3$ exhibit one endothermic peak with onset near to 968\,K. This peak is due to the L $\rightleftharpoons$ LiF+LiLuF$_4$ eutectic, and the 20\,mol\%\,LuF$_3$ DSC curve shows a single peak at this temperature, since the eutectic point is very close to this composition. The second broader endothermic peaks at 17 and 27\,mol\%\,LuF$_3$ curves mark the end of melting of the primary phase (LiF and LiLuF$_4$ respectively) at the liquidus.

Curves with compositions of 50 and 58 \,mol\%\, illustrate the LLF congruent melting at 1126\,K, and the second eutectic reaction at 1115\,K, respectively.  DSC curve for 80\,mol\%\,LuF$_3$ shows, beyond the small peak representing the liquidus line, also two intermediate peaks with onset at 1115 \,K and 1216\,K corresponding to the second eutectic reaction and the LuF$_3$ polymorphic transition from orthorrombic to hexagonal, respectively.

Finally, in the LuF$_3$ curve the polymorphic transition (1216\,K) and fusion (1403\,K) can be recognized very clearly. The main features of the LiF--LuF$_3$ experimental phase diagram are summarized in Table \ref{table02} together with some previously reported data. In general the data obtained in this work are in agreement with those from Harris \textit{et al.} phase diagram \cite{Harris83}.

\begin{table*}[htb]
\caption{Main features of LiF$-$LuF$_3$ experimental phase diagram and the previously reported data.}
	  \begin{minipage}{15cm}
    \renewcommand{\arraystretch}{1.0}
\begin{tabular}{llcrclcr}
\hline
                                        & \multicolumn{3}{l}{Composition (Mol \%\,LuF$_3$)} &  & \multicolumn{3}{l}{Temperature (K)} \\
 \cline{2-4}
 \cline{6-8}
 Equilibrium Reaction 
                                                          & This work & Harris \cite{Harris83} & Thoma \cite{thoma01} &  & This work  & Harris \cite{Harris83} & Thoma \cite{thoma01}  \\
\hline 
{\small L $\rightleftharpoons$ LiF+LLF}                    & 20     & 20   &   22   &   &  968   & 977   &   968    \\
\small (Eutectic) \\
{\small L $\rightleftharpoons$ LLF}                        & 50     & 50   &   50   &   &  1126  & 1123  &  1098    \\
\small (Congruent melting) \\                                    
{\small L $\rightleftharpoons$ LLF+LuF$_3 (S1)$}          & 58     & 58   &    54    &   &  1115   & 1105  &  1083    \\
\small (Eutectic) \\
{\small L+LuF$_3 (S1)$ $\rightleftharpoons$ L+LuF$_3 (S2)$} & --     & --  &  --     &   &  1216  &  --    &   1218   \\
\small (Polymorphic transition) \\
\hline
\end{tabular}
\end{minipage}
\label{table02}
\end{table*}

\begin{figure}[htb]
\centering
\includegraphics[width=0.47\textwidth]{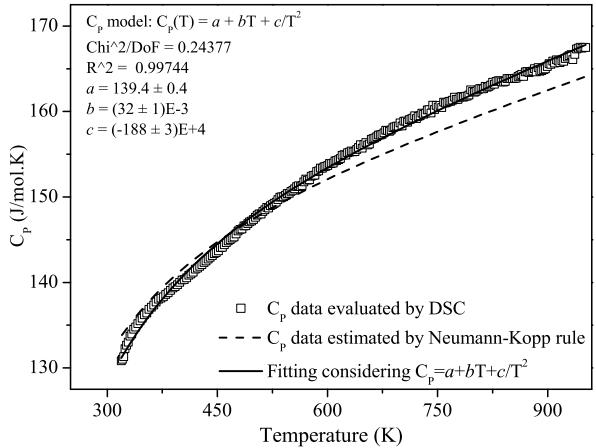}
\caption{LiLuF$_4$ $C_P$ data measured experimentally through DSC technique and $C_P$ function estimated by Neumann-Kopp rule.}
\label{fig2}
\end{figure}

The heat of fusion and polymorphic transition ($\Delta H_{PT}$) for LiF, LuF$_3$ and LiLuF$_4$ were calculated from the DSC peak area of these thermal effects for each compound. In a previous work, the experimental $\Delta H_{PT}$ for LiF was reported and reasonable agreement with the literature value was found \cite{Barin93, iasantos01}. Compared to the literature data, the main difference in the results were noticed for LuF$_3$ thermal events. The evaluated $\Delta H_{PT}$ for LuF$_3$ in this work are approximately 20\,\% smaller than those reported by Lyapunov \textit{et al.} \cite{Lyapunov04}. Since our DSC experiments were repeatedly confirmed and the raw materials used were very pure, these new results have been taken into account to the final calculation of the LiF--LuF$_3$ phase diagram.
The heat of fusion and heat capacity ($C_P$ function) for the LiLuF$_4$ compound have not yet been reported in the literature. Therefore, for the first time  these calorimetric properties were evaluated. Figure \ref{fig2} compares the $C_P$ data calculated through DSC technique and those estimated using the corresponding $C_P(T)$ functions of end members compounds and considering the Neumann-Kopp rule. One can see, for the considered temperature range, there are no signicant deviations (not bigger than 3\%) between the experimental and estimated $C_P$ data. Furthermore, taking into account the polynomial fitting of the DSC experimental data using the function (\ref{eq1}), the $a$, $b$ and $c$ parameters obtained define the $C_P(T)$ function for the LiLuF$_4$ compound.

\begin{figure}[htb]
\centering
\includegraphics[width=0.47\textwidth]{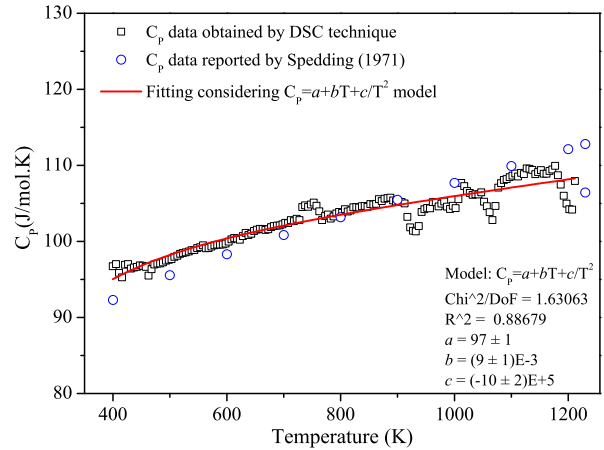}
\caption{LuF$_3$ $C_P$ data obtained experimentally through DSC technique and $C_P$ data reported previously by Sppeding \cite{Spedding71}.}
\label{fig3}
\end{figure}

Heat capacity re-evaluation for the LuF$_3$ orthorhombic phase was also carried out by DSC (Figure \ref{fig3}). The results are in reasonable agreement with those reported previously by Spedding \textit{et al.} \cite{Spedding71}. Some small deviations from the general behavior of curve were observed for the measured $C_P$ data (\textit{e.g.} 930 and 1065\,K). Those undesirable effects are due small fluctuations on the DSC base line and they can be disregarded. Taking that aside, the difference between the evaluated and reported data is not larger than 4\%, even for the higher temperatures considered. Therefore, the $C_P(T)$ parameters set by fitting a polynomial model (\ref{eq1}) on DSC experimental data were assumed to perform the numerical optimization in the LiF--LuF$_3$ system. 

\subsection{Thermodynamic assessment}

Once the $T-X$ experimental points are available and the calorimetric data for each compound belonging to the system were evaluated or collected from the literature, a proper optimization of the binary LiF--LuF$_3$ phase diagram became possible. The excess Gibbs energy for liquid phase has been optimized according to the Redlich-Kister polynomial model using the Bayesian Optimization Algorithm of FactSage \cite{FactSage6_2}. 

\begin{figure}[htb]
\centering
\includegraphics[width=0.45\textwidth]{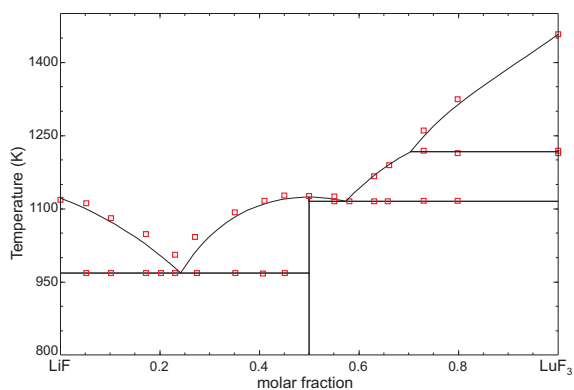}
\caption{Assessed phase diagram of the system LiF$-$LuF$_3$ (lines) together with the experimental data from DSC (squares).}
\label{fig4}
\end{figure}

Figure~\ref{fig4} presents the LiF$-$LuF$_3$ optimized phase diagram together with the $T - X$ experimental points for comparison. It may be highlighted an excellent agreement of the assessed temperature values for the eutectic reactions, polymorphic transition and even for the liquidus line. LiLuF$_4$ melting point was confirmed at $\sim$1123\,K. The value calculated for LiF/LiLuF$_4$ eutectic composition was about 3\,mol\% larger than the experimental value. However, it should be noticed that, experimentally is pretty hard to separate close thermal events, as it happens with liquidus and solidus line nearby to the eutectic composition. In general, the experimental determination of the liquidus from offsets is much less accurate than the determination of eutectic or phase transition temperatures. The assessed enthalpy and entropy at 298.15\,K temperature for the LuF$_3$ (orthorhombic and hexagonal phases) and for LiLuF$_4$, are listed in table~\ref{table01}. The excess parameters calculated, given by $L_0$ and $L_1$ in the Redlich-Kister model (equation~\ref{eq5}) were $L_0=-69089.24+49.37T$ and $L_1=8903.63-1.58T$. Even though those numbers can not be taken as unique, they represent the best assessment achieved. Together with the other data presented in this work, that numerical simulation gives a complete thermodynamic description for the LiF--LuF$_3$ binary phase diagram.

\section{Conclusions}

Thermodynamic assessment has been performed on the LiF$-$LuF$_3$ binary system and the excess Gibbs energy terms for the liquid phase could be properly optimized using the Redlich-Kister polynomial model. Based on our experimental results and the thermodynamic assessment it was possible to confirm the values for the two eutectic temperatures within this system. A $C_P(T)$ function of LiLuF$_4$ compound has been set for the first time and it was in accord with Neumann-Kopp rule. It has been successfully confirmed the previously $C_P$ data found in the literature  for LuF$_3$ (orthorhombic phase). $\Delta H${(298.15\,K)} and $S${(298.15\,K) have been assessed for LuF$_3$ and LiLuF$_4$, and a re-evaluation for $\Delta H_{PT}$ data in the LuF$_3$ compound was performed. This work offers a more complete thermodynamic description of LiF$-$LuF$_3$ system and makes it possible further thermodynamic assessment for ternary and multi component phase diagram based on this binary one.

\section*{Acknowledgments}

The authors acknowledge financial support from CAPES (Grant no. 368/11) and DAAD (Grant no. po-50752632) in the framework of the PROBRAL program, and from CNPq (477595/2008-1). One of the authors (I.A. dos Santos) acknowledges financial support from DAAD-CAPES-CNPq (290111/2010-2). This work also was supported by the EU Commission in the Seventh Framework Programme through the ENSEMBLE project (Grant Agreement Number NMP4-SL-2008-213669).











\end{document}